\documentclass[twocolumn]{aastex63}

\definecolor{blue-violet}{rgb}{0.54, 0.17, 0.89}

\usepackage{multirow}
\usepackage{makecell}

\graphicspath{{./}{figures/}}

\accepted{10th July 2022 to ApJ} 

\shorttitle{First Infallers}
\shortauthors{Smith et al.}

\begin{document}

\title{The First Fall is the Hardest: The Importance of Peculiar Galaxy Dynamics at Infall Time for Tidal Stripping Acting at the Centers of Groups and Clusters}

\correspondingauthor{Rory Smith}
\email{rory.smith@kasi.re.kr}

\author{Rory Smith}
\affiliation{Departamento de Física
Universidad Técnica Federico Santa María
Vicuña Mackenna 3939, San Joaquín, Santiago de Chile}

\author{Paula Calder\'on-Castillo}
\affiliation{Departamento de Astronom\'ia, Universidad de Concepci\'on, Casilla
 160-C, Concepci\'on, Chile}

\author{Jihye Shin}
\affiliation{Korea Astronomy and Space Science Institute (KASI), 776 Daedeokdae-ro, Yuseong-gu, Daejeon 34055, Korea}

\author{Mojtaba Raouf}
\affiliation{Leiden Observatory, Leiden University, Niels Bohrweg 2, 2333 CA Leiden, the Netherlands}

\author{Jongwan Ko}
\affiliation{Korea Astronomy and Space Science Institute (KASI), 776 Daedeokdae-ro, Yuseong-gu, Daejeon 34055, Korea}
\affiliation{University of Science and Technology (UST), Daejeon 34113, Korea}



\begin{abstract}
Using dark matter only N-body cosmological simulations, we measure the pericentre distance of dark matter halos on their first infall into group and cluster halos. We find that the pericentre distance (R$_{\rm{peri}}$) is an important parameter as it significantly affects the strength of tidal mass loss in dense environments, and likely other environmental mechanisms as well. We examine what determines the R$_{\rm{peri}}$ value and find that, for most infallers, the dominant parameter is V$_{\rm{\perp}}$, the tangential component of the orbital velocity as the halo enters the group/cluster halo for the first time. This means that the strength of tidal stripping acting near the cores of groups/clusters are strongly influenced by the external peculiar velocity field of the large scale structure surrounding them, which differs between clusters, and is sensitive to the mass ratio of infaller to host. We find that filament feeding also partially contributes to feeding in low V$_{\rm{\perp}}$ halos. Dynamical friction can also play a role in reducing R$_{\rm{peri}}$ but this is only significant for those few relatively massive infallers ($>$10\% of the mass of their host). These results highlight how the response of galaxies to dense environments will sensitively depend on dynamics inherited from far outside those dense environments.
\end{abstract}


\section{Introduction} \label{sec:intro}

There is now an abundance of evidence for physical mechanisms operating in dense environments such as galaxy clusters that can transform the properties of galaxies that fall into them. The morphology-density relation \citep{Dressler1980, Whitmore1993}, the star-formation--density relation \citep{Kennicutt1983}, and the atomic gas--density relation \citep{Haynes1984} all indicate that living in dense environments has consequences for galaxy properties. 

A variety of physical mechanisms that could cause these changes have been proposed. The large mass of the cluster halo creates a deep, extended potential well whose tidal field can strip dark matter, stars and gas from satellites \citep{Merritt1984}. Meanwhile, within the cluster there are numerous other cluster members that can have fly-by encounters with each other, a process known as `harassment' \citep{Moore1996,Moore1999,Gnedin2003a,Smith2010a,Smith2013a,Bialas2015, Smith2015,smith2021}. Besides tidal stripping mechanisms there are hydrodynamical mechanisms, such as ram pressure stripping \citep{GunnGott1972}, which can remove the gas content of a galaxy but leave the stellar content largely unaffected. There is increasing evidence that no single mechanism dominates, and that a variety of mechanisms may conspire to remove the cold gas content and eventually halt star formation (\citealp{Marasco2016}, see also \citealp{Cortese2021} for a recent review).


A satellite galaxy's orbit within its hosting cluster or group is expected to significantly influence the efficiency of the physical mechanisms operating in dense environments. A more plunging orbit will take a galaxy deeper into the host to radii where the environmental density is higher. From a gravitational point of view, this is expected to result in stronger tidal mass loss. First, the strength of the tidal field from the main cluster halo increases near the cluster centre \citep{Byrd1990}. Second, near the cluster centre there are more cluster members and so there is an increased chance of a galaxy suffering high speed encounters \citep{Smith2010a}. Third, as we will show in this study, a galaxy with a plunging orbit also tends to have a higher orbital velocity at the pericentre of its orbit (the closest approach of the galaxy to its host's center), causing the tidal field to evolve very rapidly, perhaps so fast that the galaxy cannot respond in time. As a result, such a galaxy responds impulsively to the rapid change of tidal field, much like in a high speed encounter between cluster members \citep{Gonzalez2005}. 

A small pericentre is also expected to influence the response of hydrodynamical gas mass loss mechanisms as well. The strength of ram pressure stripping is expected to depend on two main terms -- the density of the intracluster medium (ICM) and the square of the velocity at which the galaxy passes through that medium \citep{GunnGott1972}. With a smaller pericentre, a galaxy will tend to pass through a denser ICM, and have a higher peak orbital velocity causing the ram pressure to be significantly stronger from the radial dependency of both terms combined \citep{Vollmer2001b}. Similar to with gravitational tidal fields, if the pericentre is small the ram pressure may increase and decrease very rapidly. This may occur so quickly that gas which is technically unbound by the ram pressure does not have time to be fully removed from the disk by the time that the ram pressure has decreased following the pericentre passage \citep{Roediger2007}. Such gas may then fall back onto the disk at later times under the action of gravity \citep{Vollmer2001, Schulz2001, Kapferer2009, Quilis2017}



Given the important role that R$_{\rm{peri}}$ plays in dictating the efficiency of environmental mechanisms in dense environments, it is important to understand what factors dictate R$_{\rm{peri}}$ and these can be studied using simulations. There are many papers in the literature devoted to the study of the orbital parameters of infalling dark matter halos in cosmological simulations (e.g., \citealt{Tormen1997,Vitvitska2002,Benson2005,Zentner2005,Wang2005,Khochfar2006,Libeskind2014,Shi2015,Jiang2015,Li2020,Benson2021}). Their results tend to support a picture in which the pericentres are expected to be smaller in higher mass hosts, or when infalling halos are more massive compared to their hosts, because the tangential component of their orbital motion when crossing into the host is lower. They also tend to find there is only a weak dependency on redshift, at least since redshift one \citep{Wetzel2011,Li2020,Benson2021}. However, the majority of these studies rely on identifying the infallers at the moment they fall into the host and recording their orbital dynamics at that time, then extrapolating their future orbit and pericentre. This is typically accomplished assuming point masses for the host and infaller halo, or a fixed and often spherical potential for the cluster. Of course, this neglects the fact that the host halos are extended potentials with typically non-spherical and complex shapes that grow and evolve with time \citep{Benson2010}. Furthermore, dynamical friction is generally neglected or approximated, and in most cases the resulting mass evolution of the satellite is not measured directly within the simulation.

In this paper, we attempt to overcome these limitations by directly measuring the R$_{\rm{peri}}$ of halos infalling into groups and clusters in a cosmological simulation. Rather than approximating their orbits inside the host, we self-consistently track the orbit of the infaller halo from before entering the cluster until passing first pericentre, using the merger tree of our DM-only cosmological simulation. As a result, our R$_{\rm{peri}}$ measurements naturally include the effect of the extended density distribution, shape and time evolution of the cluster potential, and dynamical friction naturally arises between the infaller halo and its host. By measuring the evolving mass of the halo as we track it throughout its first infall, we can also directly measure how R$_{\rm{peri}}$ impacts on the strength of tidal mass loss. Finally, because some halos will enter the cluster as members of groups in our cosmological simulation, their tangential velocities may be increased by the peculiar motions within the group \citep{Li2020} potentially increasing R$_{\rm{peri}}$. On the other hand, the stronger dynamical friction that a massive group experiences could also tend to drag group members that remain bound to their group down to smaller R$_{\rm{peri}}$ values \citep{Choque2019} which, thanks to our approach, is an effect that will be captured and could influence the resulting R$_{\rm{peri}}$ values we measure.

Using our simulations, we are able to study how the peculiar motions of infaller halos that are inherited from large scale dynamics outside of the cluster (i.e., an external effect), combined with dynamical friction operating inside the cluster (i.e., an internal effect) combine to dictate the R$_{\rm{peri}}$ value, and we will show how this, in turn, impacts on the tidal mass loss suffered by the halo.

\begin{figure*}[ht]
\begin{center}
\includegraphics[width=19.0cm]{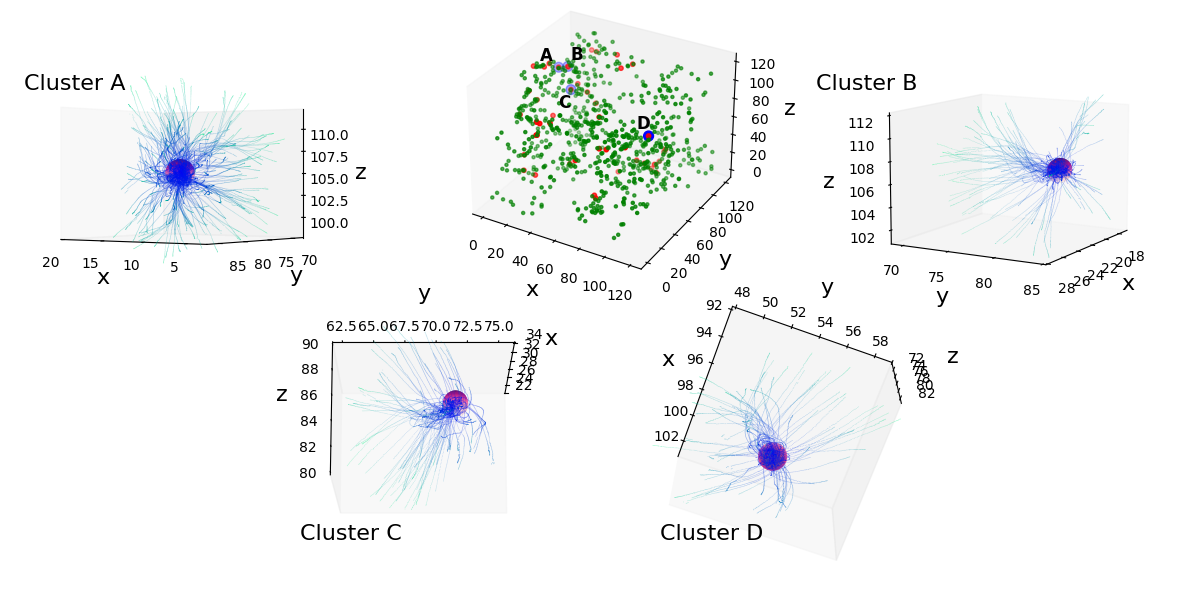}
\end{center}
\caption{Central panel: 120 Mpc/h cosmological box of clusters (red symbols) and groups (green symbols). Blue points indicate zoomed-in clusters (labelled `A'--`D'). Surrounding panels: Zoom-in of four representative clusters (labelled `Cluster A--D') with a deep-pink sphere indicating their $z$=0 position (size equal to the cluster R$_{200}$). Blue lines indicate the orbital trajectories for all their infallers, coloured darker blue according to proximity to the final cluster position.}
\label{fig:120Mpcbox}
\end{figure*}

Our paper has the following structure; in Section 2 we describe the simulations and methodology, in Section 3 we present our results, and in Section 4 we summarise and conclude.

\section{Method}
\label{sec:method}

For this study, we analyse a dark matter-only cosmological simulation. Our cosmological initial conditions were built for an initial redshift=200, using the Multi Scale Initial Condition software (MUSIC; \citealt{Hahn2011}). The simulated volume is a cube with a 120~Mpc/h side-length. Each dark matter particle has a mass of  $1.072\times10^9$~M$_{\odot}/h$.
We evolve the initial conditions down to redshift=0, using GADGET-3 \citep{Springel2001}. The following cosmological parameters are assumed; $\Omega_m = 0.3, \Omega_{\Lambda} = 0.7, \Omega_b = 0.047$ and $h_0 = 0.684$. We output snapshots $\sim$100~Myr apart. These frequent outputs are useful for building merger trees and accurately tracing out the orbits of halos. With this output frequency, we can typically recover the true pericentre to within a few kiloparsecs (see Appendix Figure \ref{fig:missedperis}).

We run the ROCKSTAR halo finder to identify halos \citep{Behroozi2013}. This uses a 6 dimensional phase space hierarchical Friends-of-Friends (FoF) algorithm to better identify satellites in host halos than is possible with a 3 dimensional FoF approach. We include halos down to 20 dark matter (DM) particles, meaning our minimum halo mass is $2\times10^{10}$~M$\odot$/h. However, we take a cut for halos whose infall mass is less than $1\times10^{11}$~M$\odot$/h to ensure we can track the tidal mass loss of halos down to $\sim$20\% of their infall mass. The peak of the density is used to define the center of a halo. The merger tree is built with {\sc{Consistent Trees}} \citep{Behroozi2013}, which combines particle IDs with halo trajectory information to improve the linking of halos between snapshots.

\begin{figure}[ht]
\begin{center}
\includegraphics[width=8.5cm]{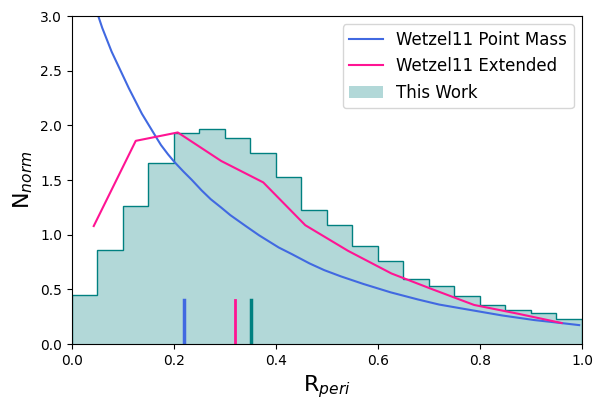}
\end{center}
\caption{The main R$_{\rm{peri}}$ distribution (teal filled) for all the infaller sample. The lines are from \cite{Wetzel2011}. The blue curve is the result when the cluster potential was treated as a point mass. The magenta curve is result when an analytical extended host potential was used. The small vertical lines along the x-axis indicate the median value of the three distributions in the legend.}
\label{fig:mainrperidistrib}
\end{figure}

At $z$=0, there are 44 central halos with cluster masses (M$_{200}>10^{14}$~M$_\odot$/h) and 692 central halos with group masses ($10^{13}<$ M$_{200}$(M$_\odot$/h)$<10^{14}$) in the full 120~Mpc/h simulation volume. We refer to these cluster and group halos as the `host halos'. The spatial distribution of the host halos within the 120 Mpc/h cosmological box is shown in the centre panel of Fig. \ref{fig:120Mpcbox}. We track the trajectories of all halos that fall into these hosts (i.e., cross the R$_{200}$ of the host halo, measured at the time of infall), which gives us a sample of 65,640 halos from which we further subsample. The blue lines show the trajectories of all the infaller halos in four representative clusters as shown in the sub-panels of Fig. \ref{fig:120Mpcbox}. The final $z=0$ position of the cluster halo is shown as a deep-pink sphere, and it is clear that the infaller halo trajectories are highly non-isotropic, with a preference for particular infall directions as a result of filament feeding. In this study, we limit ourselves to consider only infaller halos that have at least one pericentre, which gives us a sample of 42,353 halos, that we refer to as the `infaller halos'. Some of these survive until $z=0$ ($\sim$16,000 halos) and the rest are destroyed inside their host halos. If we define the host by their $z=0$ mass, $\sim$11,000 of all the infallers fall into clusters (M$_{200}$($z$=0)$>10^{14}$~M$_\odot$), and $\sim$31,500 of the infallers fall into groups (M$_{200}$($z$=0)$=10^{13}-10^{14}$~M$_\odot$).

We measure a number of properties of the infaller halos at the instant that they cross their host halo's R$_{200}$ for the first time. These include; infaller halo mass (M$_{\rm{inf}}$), host mass (M$_{\rm{host}}$), the mass ratio of the two (M$_{\rm{rat}}$=M$_{\rm{inf}}/$M$_{\rm{host}}$), 
and the time at which the halo entered the host (t$_{\rm{inf}}$). To measure the strength of peculiar motions of infallers, we also measure the perpendicular component of the orbital velocity normalised by the host velocity dispersion (V$_\perp$), also measured at the moment of infall. Finally, we also test if the infaller halo is falling into the host in isolation or as part of a larger substructure such as a group and, if so, we record the mass of the group.

After the halos have entered the host, we track the evolution of the halo's orbit until they reach their first pericentre (at t$_{\rm{peri}}$), when we record the normalised pericentre distance (R$_{\rm{peri}}$) as the minimum 3-dimensional separation between the infaller halo's centre and the host's centre, normalised by the host's R$_{200}$.  We also track the mass evolution of the infaller halo, normalised by its mass at the time of infall.

\begin{figure*}[ht]
\begin{center}
\includegraphics[width=17cm]{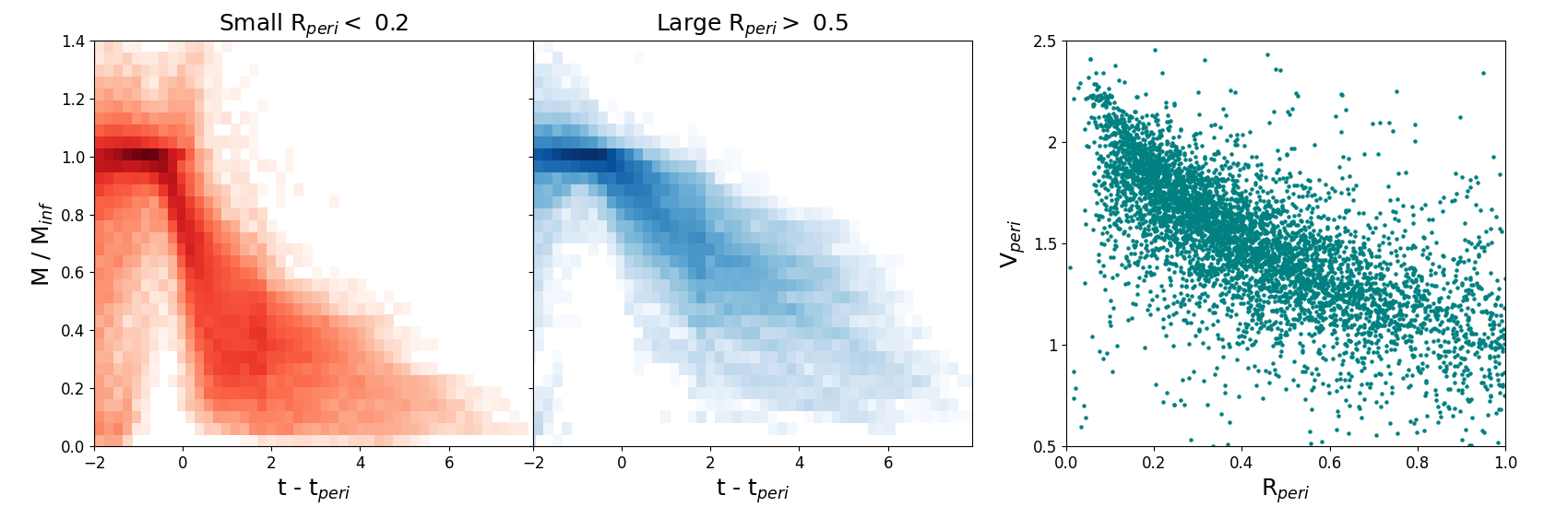}
\end{center}
\caption{Tidal mass loss of infaller halos with small pericentres (R$_{\rm{peri}}<0.2$; left panel) and large pericentres (R$_{\rm{peri}}>0.5$; center panel). The y-axis is the halo mass of the infaller normalised by its mass at the time of infall. The x-axis is time since pericentre passage in Gyr, with negative x-values being before the infaller reaches pericentre. Halos with small R$_{\rm{peri}}$ values suffer stronger and more rapid tidal mass loss. Right panel: Orbital velocity at pericentre normalised by host velocity dispersion (V$_{\rm{peri}}$) versus R$_{\rm{peri}}$. Halos with smaller pericentres tend to have higher normalised pericentre velocities.}
\label{fig:tidalmassloss}
\end{figure*}

\section{Results}
\label{sec:results}

\subsection{The total Rperi distribution}

We begin by looking at the total distribution of R$_{\rm{peri}}$ for all of the sample of infallers that reach pericentre, as shown by the teal filled histogram in Fig. \ref{fig:mainrperidistrib}. The distribution peaks at R$_{\rm{peri}}=0.26$ (N.b., R$_{\rm{peri}}$ is normalised by the host R$_{\rm{200}}$ at the time of pericentre). The first quartile, median and third quartile of the distribution is R$_{\rm{peri}}$=0.22, 0.35, and 0.52, respectively (see small vertical lines on x-axis). We compare our distribution to the total R$_{\rm{peri}}$ distribution from \cite{Wetzel2011} (the blue and magenta line). The blue line shows the results obtained when the cluster potential was treated as a point mass. The magenta line shows the result when instead an extended analytical potential was applied for the cluster potential. Clearly the extended analytical potential for the cluster is a closer match to our result as both histograms have a peak and the numbers fall as R$_{\rm{peri}}$ approaches zero, unlike in the point-mass treatment of the cluster. However, there are still differences, with our distribution preferring slightly larger values of R$_{\rm{peri}}$, perhaps because the our infallers orbit in the true cluster potential rather than being approximated by an analytical potential.

\subsection{Rperi controls Tidal Mass loss}

In Fig. \ref{fig:tidalmassloss}, we illustrate the importance of the R$_{\rm{peri}}$ value for tidal mass loss of infallers after entering the host. In the plots, we stack the individual mass evolution tracks of infaller halos from the moment of their infall into the host halo. Therefore, a single halo contributes multiple data points to the plot in the form of a track, and the colour is darker where multiple tracks overlap. The $y$-axis is normalised by the mass at infall time, thus all halos start their track when they first enter the host at $y=1$. We plot t-t$_{\rm{peri}}$ on the x-axis, meaning the $x$-axis is 0 at the moment of pericentre. This is chosen for a reason. As can be seen, initially most halos lie on the dark, nearly horizontal strip until they reach pericentre. This means typically they do not lose much mass throughout their first infall until they are close to pericentre, and then mass loss begins suddenly. Similar results can also be found in \cite{Knebe2011,Behroozi2014}. Thus, by plotting t-t$_{\rm{peri}}$ on the x-axis, we better align the mass evolution tracks in the stack. 

We split the infaller sample into a `small Rperi' (R$_{\rm{peri}}<0.2$; left panel) and `large Rperi' (R$_{\rm{peri}}>0.5$; center panel) subsample, in order to compare their mass evolution. The limiting values of R$_{\rm{peri}}$ for these two subsamples are chosen to give two extremes, in order to better illustrate the R$_{\rm{peri}}$ dependency. By eye, the mass evolution appears sensitive to the R$_{\rm{peri}}$ value, with smaller R$_{\rm{peri}}$ producing stronger and more rapid tidal mass loss. We quantified this result by measuring the percentage of infaller halos that suffer strong tidal mass (defined as M/M$_{inf}<0.4$) by 2~Gyr after pericentre. For R$_{\rm{peri}}<0.2$, 60.2\% suffer strong tidal mass loss compared to only 13.1\% for R$_{\rm{peri}}>0.5$. Physically, this is simple to understand as smaller pericentres take the infaller deeper into the host's potential where tides are stronger, and where more high speed encounters with cluster members occur \citep{Smith2010a}. 

However, we note that the efficiency of non-gravitational environmental effects are likely also a sensitive function of pericentre distance. Ram pressure strength depends on the density of the intracluster medium (ICM) times the galaxy velocity through that ICM \citep{GunnGott1972}. Although we do not model ram pressure stripping directly here (there are dark matter only simulations), we can expect that halos with smaller pericentres would reach higher peak densities of the ICM. Also, as shown in the right panel of Fig. \ref{fig:tidalmassloss}, orbits with smaller pericentres result in higher peak velocities. Thus, a combination of denser ICM and higher orbital velocities for halos with smaller pericentres should cause stronger peak ram pressures \citep{Roediger2007}. 

\subsection{What dictates R$_{\rm{peri}}$?}

\begin{figure*}[ht]
\begin{center}
\includegraphics[width=8.5cm]{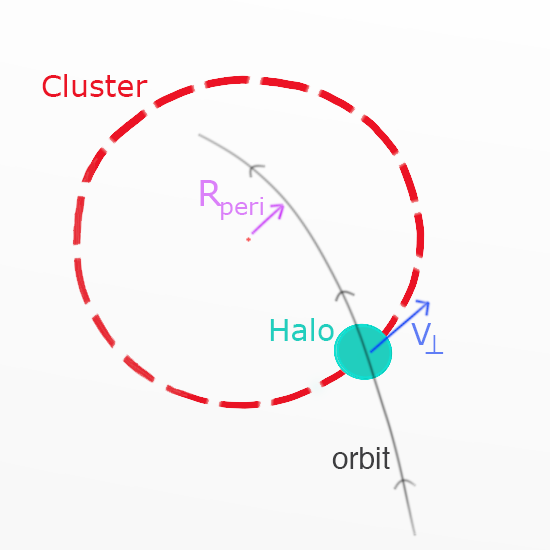}
\includegraphics[width=8.5cm]{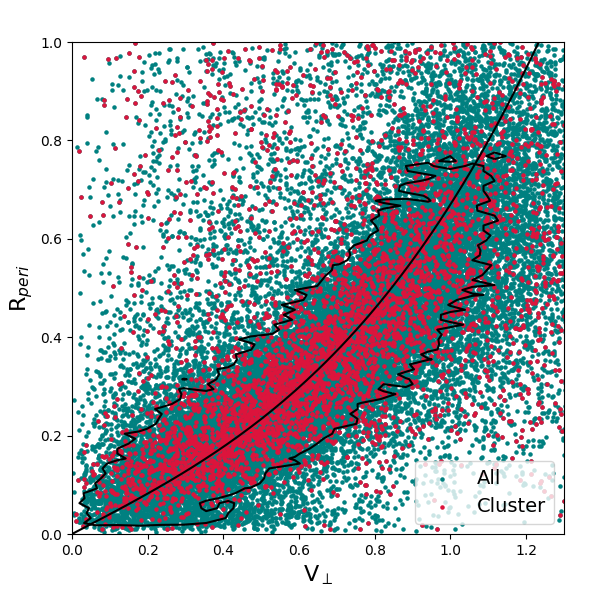}
\end{center}
\caption{Left: Cartoon schematic illustrating how increasing V$_\perp$ results in increased R$_{\rm{peri}}$. The red dashed circle is the cluster R$_{200}$. The cyan filled circle is an infaller halo at the moment of entering the cluster. The blue arrow labelled `V$_\perp$' shows the perpendicular component of the halos orbital velocity. Right: R$_{\rm{peri}}$ vs V$_\perp$ plot (normalised axes) for the full sample (teal) and for cluster-mass hosts only (red). There is a clear trend, which is independent of host mass. The smooth black curve is an approximate fit to the trend (R$_{\rm{peri}}$=0.27 V$_\perp^3$+0.40 V$_\perp$) which we apply in Section \ref{sec:Infalltime_MassRatio}. The scatter about the trend is small. The black contour contains 95\% of the `All' sample, meaning most objects are found within $\pm0.1$ of the fit line.}
\label{fig:RperiVperp}
\end{figure*}

\begin{figure*}[ht]
\begin{center}
\includegraphics[width=8.5cm]{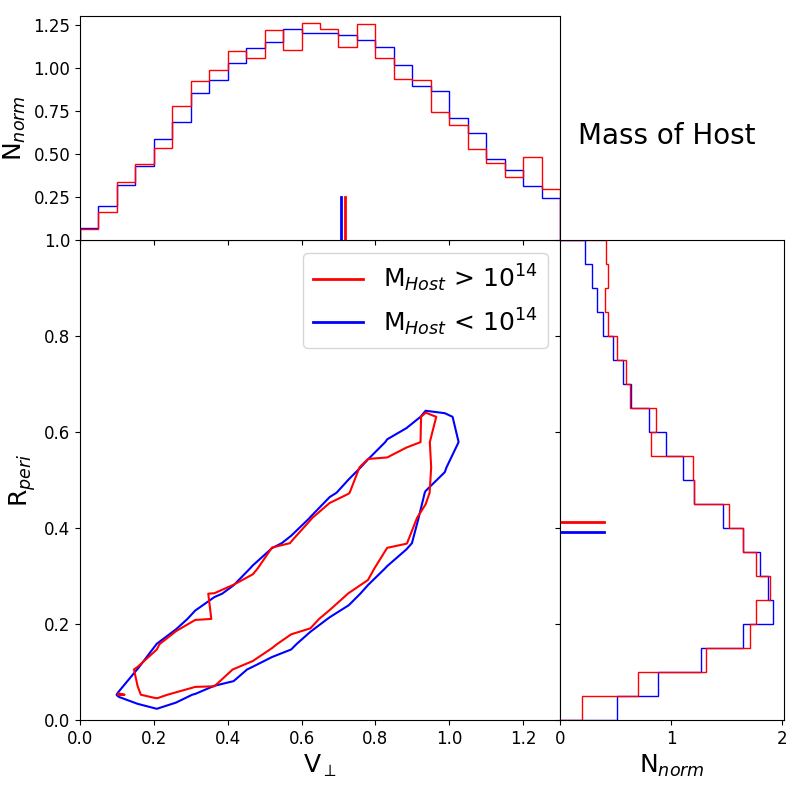}
\includegraphics[width=8.5cm]{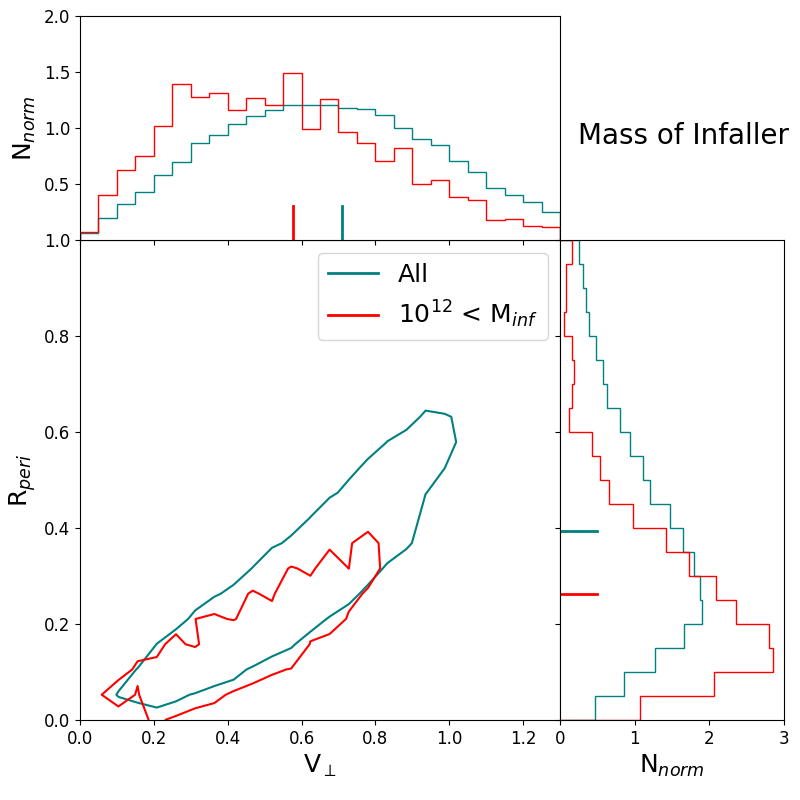}
\end{center}
\begin{center}
\includegraphics[width=8.5cm]{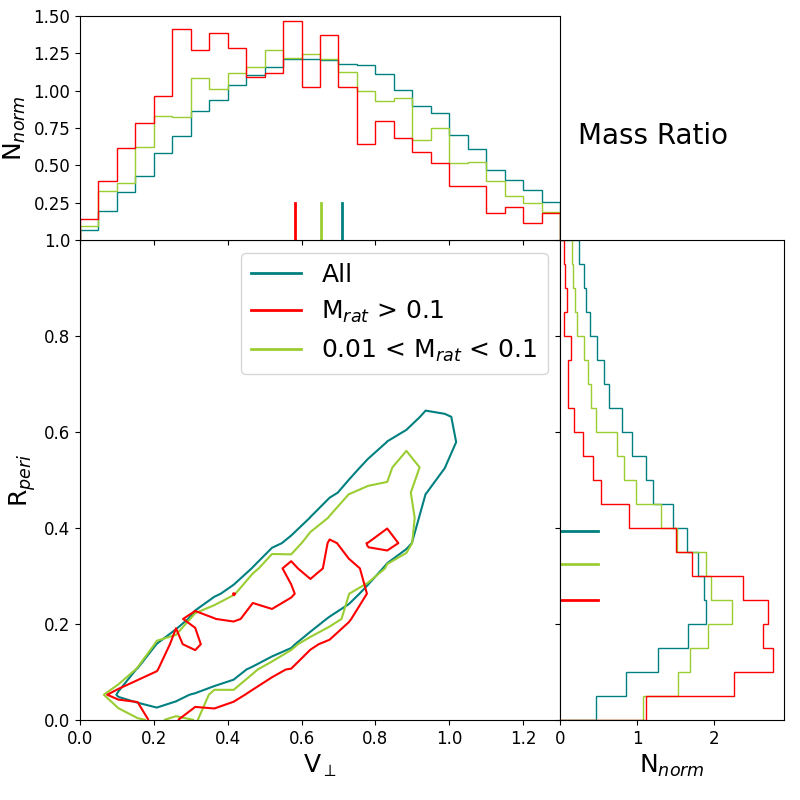}
\includegraphics[width=8.5cm]{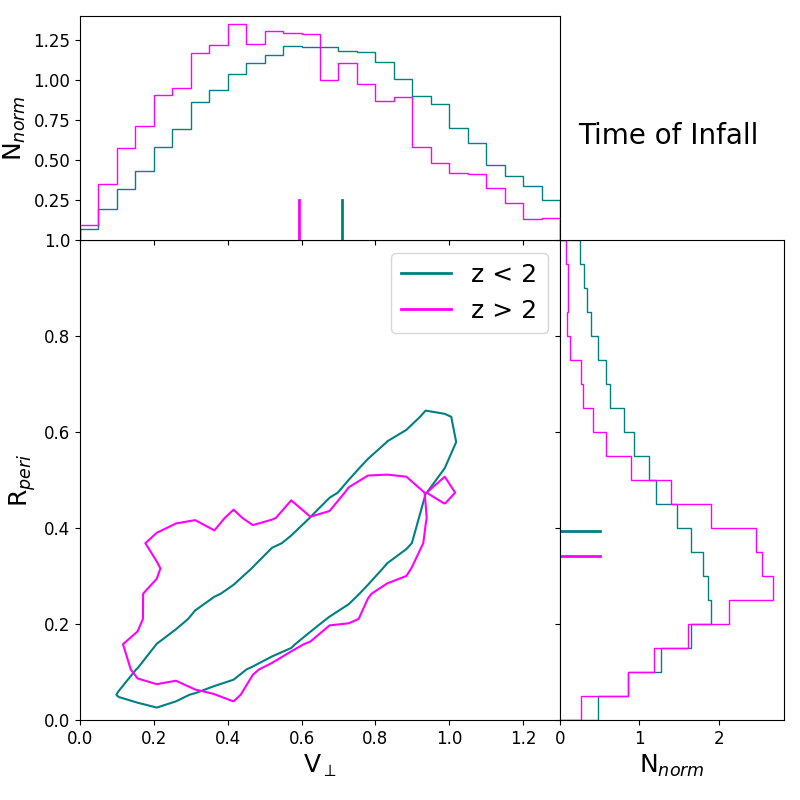}
\end{center}
\caption{Plots illustrating the relation between R$_{\rm{peri}}$ and V$_\perp$. The main panel is R$_{\rm{peri}}$ vs V$_\perp$, and the contour encompasses 68\% of the data points. The upper-histograms shows the normalised V$_\perp$ distribution (a vertical integration of the R$_{\rm{peri}}$ vs V$_\perp$ plot) and the side panel shows the normalised R$_{\rm{peri}}$ distribution (a horizontal integration of the R$_{\rm{peri}}$ vs V$_\perp$ plot). We subsample by various parameters; upper-left we split by host mass, upper-right by mass of the infaller halo, bottom-left by the mass ratio of the infaller to the host, and lower-right by the time of infall (see legend for line colours). A short line on the histogram axis indicates the median value for each distribution.} 
\label{fig:rperibecauseofvperp}
\end{figure*}

\begin{figure*}[ht], 
\begin{center}
\includegraphics[width=17cm]{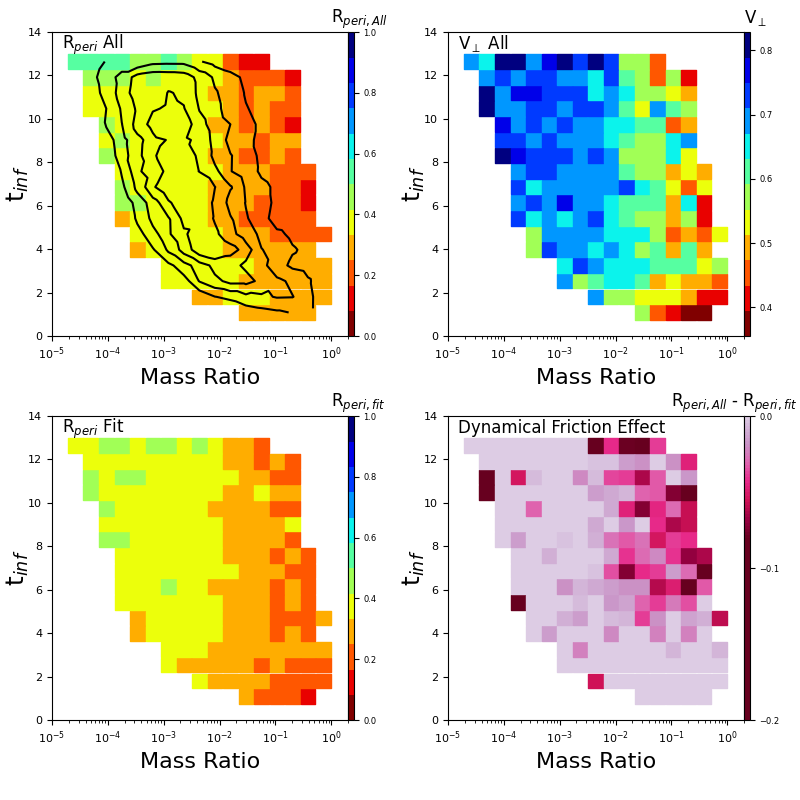}
\end{center}
\caption{Age of the Universe in Gyr at the time of infall (t$_{inf}$) versus mass ratio plots. Top left: Pixels are coloured by the median R$_{\rm{peri}}$ of halos that fall within them. Black contours indicate 100, 200, 300, and 400 objects in a pixel. Top right: Pixels are coloured by median V$_\perp$. Bottom left: Pixels are coloured by the R$_{\rm{peri}}$ value that would be expected if dynamical friction were heavily suppressed (see text for details). R$_{\rm{peri}}$ values are based only on the median V$_\perp$ value, using the fit line from the right panel of Fig. \ref{fig:RperiVperp} (see text for details). Bottom right: Pixel-by-pixel subtraction of the upper-left and lower-left panels to show how much the measured R$_{\rm{peri}}$ values reduce compared to those when dynamical friction is heavily suppressed in the sample. Pixels are blank if they contain less than 20 halos.}
\label{fig:grids}
\end{figure*}

As the pericentre distance has proven important for the strength of environmental mechanisms acting in clusters and groups, we now turn our attention to what processes dictate the value of R$_{\rm{peri}}$. Physically, these can be broadly split into external processes (e.g., the peculiar motions of halos, inherited from their dynamics within the large scale structure beyond the cluster) and internal processes (e.g., dynamical friction) operating within the cluster.

\subsubsection{Peculiar Motions versus Dynamical Friction}

When an infaller halo crosses R$_{200}$ of the host halo for the first time, we measure the perpendicular component of its velocity with respect to a radial line between the host centre and the infaller. We normalise the velocity by the host halo's velocity dispersion. The normalised perpendicular velocity is referred to as V$_{\rm{\perp}}$.

V$_{\rm{\perp}}$ is expected to influence R$_{\rm{peri}}$, as illustrated in the cartoon schematic in the left panel of Fig. \ref{fig:RperiVperp}. Higher V$_{\rm{\perp}}$ values should result in increased R$_{\rm{peri}}$. 

In the right panel of Fig. \ref{fig:RperiVperp}, we show a scatter plot of R$_{\rm{peri}}$ versus V$_{\rm{\perp}}$ for all of the sample that fell into their hosts since $z$=2 (teal points; labelled `All' in the legend). There is a clear correlation between the two parameters, as predicted. We choose infallers since $z$=2 so that the trend can be more clearly seen as for very early infallers the scatter increases (as we will show below). By eye, we fit the black smooth curve to the trend (the equation of the fit is given in the figure caption) for use later in Section \ref{sec:Infalltime_MassRatio}. Although by eye the scatter in the teal points appears quite large, the black contour contains 95\% of the `All' sample. Thus, the great majority of the points are found within R$_{\rm{peri}}$=$\pm$0.1 of the fit to the trend. We emphasise that, as V$_{\rm{\perp}}$ is measured when the halo enters the cluster for the first time, this highlights the crucial role the peculiar motions of halos has on environmental effects occurring deep inside the cluster itself, motions that are inherited from the dynamics of the large scale structure, far beyond the cluster outskirts.

The red points in the right panel of Fig. \ref{fig:RperiVperp} indicate a subsample where the host halo has a cluster mass (M$_{\rm{200}}>10^{14}$~M$_\odot$) at the time of the infall. The cluster mass subsample sits on the same trend as the `All' sample, indicating that the normalisation of the $y$- and $x$-axis by the host virial radius and velocity dispersion respectively, effectively removes any dependence on host mass. Although the scatter in the trend is small (95\% of the `All' sample is found between the black contours), we tried to understand the primary cause of the scatter. We tested if the concentration of the host was a factor but did not see a clear dependency. Instead, we found that the majority of the scatter could be explained by a dependency on how much of a halo's orbital motion is radial at the moment of infall. Halos with stronger radial motions tend to have smaller R$_{\rm{peri}}$ for a fixed value of V$_{\rm{\perp}}$ and vice versa.

The importance of V$_{\rm{\perp}}$ for deciding the shape of the R$_{\rm{peri}}$ distribution is further highlighted in Fig. \ref{fig:rperibecauseofvperp}. Here we split all of the infallers that fell into hosts since $z=2$ (to reduce scatter, as we will show below) into subsamples by; host mass, infaller mass, infaller to host mass ratio, and time of infall (from upper-left to bottom-right respectively). The main panel shows R$_{\rm{peri}}$ versus V$_\perp$ with a contour containing 68\% of the data points. The upper histogram shows the corresponding V$_\perp$ distribution, and the histogram on the right shows the R$_{\rm{peri}}$ distributions, with small lines indicating the medians of the distributions.

A trend between R$_{\rm{peri}}$ and V$_\perp$ is visible in all the subsamples, although the location of the trend and, in particular, the location of halos on the trend differs between the subsamples. For host mass (upper left), the trend is very similar, and the resulting distributions are thus similar too. In the upper-right and lower-left panel, the subsamples seems to roughly follow the same trend, but they are not distributed equally along it. Higher mass infallers and higher mass ratio infallers tend to have lower V$_\perp$ values. As a result of adhering to an R$_{\rm{peri}}$--V$_\perp$ trend, they end up with lower R$_{\rm{peri}}$ values too, and thus the V$_\perp$ and R$_{\rm{peri}}$ distributions show similar dependencies on each parameter. 

In addition, a small vertical offset in the trend appear for high mass infallers, and for infallers with a mass ratio$>0.1$. This means that, at a fixed V$_\perp$, R$_{\rm{peri}}$ is reduced. Neither the high mass infallers or high mass ratio infallers had higher fractions of their motion radially therefore we ruled this out as a factor. Instead, this is a key indicator of dynamical friction which acts to remove orbital energy from the infaller and shrink the radius of the orbit. Dynamical friction becomes quite ineffective when the infaller mass is less than a few percent of the host mass \citep{BoylanKolchin2008}. Indeed, for the first infallers that we consider in this study, the dynamical friction can only act for a limited time, since the infaller halo enters the cluster until it reaches first pericentre, further reducing its impact on the infaller orbit. As a result, the reduction in R$_{\rm{peri}}$ is quite weak ($<0.1$) in both the upper-right and lower-left panel, even when we have selected quite extreme infaller masses (only 5\% of infallers have M$_{\rm{inf}}>0.1$) or mass ratios (only 4\% of infallers have M$_{\rm{rat}}>0.1$).

Finally, in the lower right panel we see that early infallers (infall redshift $z>2$) show a more scattered trend. The data points cannot scatter below R$_{\rm{peri}}=0$, thus only the scatter upwards, above the trend is visible in the contours. However, this increased scatter is only visible at early times, and for most of the age of the Universe (over the last $\sim10$~Gyr, which includes 90\% of infallers), there is a well defined sequence between R$_{\rm{peri}}$ and V$_\perp$.

Summarising, we find that in most cases, V$_\perp$ is the dominant parameter controlling the R$_{\rm{peri}}$ distribution, including its dependence on other parameters such as host mass, infaller mass, mass ratio or time of infall. 
Dynamical friction can impact R$_{\rm{peri}}$ alongside the effects from V$_\perp$, but only for the upper few percent of the most massive infallers or highest mass ratios. Even for these extreme samples the distribution of V$_\perp$ plays an important role.

\subsubsection{Infall time and Mass ratio}
\label{sec:Infalltime_MassRatio}

In the top-left panel of Fig. \ref{fig:grids}, we attempt to visualise how the median R$_{\rm{peri}}$ value (shown by the colour of a pixel) changes as a function of both infall time (y-axis, measured in terms of the age of the Universe at that moment in Gyr) and mass-ratio at that infall time (x-axis) using a pixelated grid of the median R$_{\rm{peri}}$ of all halos that fall within a pixel.

There is a general horizontal gradient for R$_{\rm{peri,All}}$ to decrease as the mass ratio increases. The fact that the gradient is predominantly horizontal indicates that there is only a weak dependency on the time of infall, at least over the last $\sim10$~Gyr, as seen in other recent studies \citep{Li2020,Benson2021}. This horizontal gradient exists even when the mass ratio is very small (mass ratios$\sim10^{-4}$ up to 0.01) when dynamical friction is expected to be negligible. In this low mass ratio regime, where the majority of the halos are located (see black contours), the R$_{\rm{peri,All}}$ is accompanied by a similar gradient in V$_\perp$ (see top right panel). This suggests that the R$_{\rm{peri,All}}$ gradient is actually due to a dependency of V$_\perp$ on the mass ratio, for the majority of the halos (in agreement with \citealt{Li2020}). We note that because V$_\perp$ is measured at the time of infall, this dependency on mass ratio must have been already set beyond the cluster, within the surrounding large scale structure.

To attempt to see how V$_\perp$ alone drives the dependency of R$_{\rm{peri,All}}$ on the mass ratio, we analytically calculate R$_{\rm{peri}}$ using the fit equation from the right panel of Fig. \ref{fig:RperiVperp} (shown as a black curve in that figure). By inserting the measured V$_\perp$ value in the simulation into the fit equation, we get the analytically calculated R$_{\rm{peri,fit}}$ value. We note that by doing this the scatter about the trend will be collapsed onto the median line. But in this way we ensure that the median value of the R$_{\rm{peri}}$ is conserved for a measured V$_\perp$ value. The impact of dynamical friction on this fit line can be assumed to be heavily suppressed. This is because the sample is heavily dominated by objects with mass ratios below 0.01 for which simulations have shown that dynamical friction has a negligible impact on their first pericenter distance \citep{Lacey1993,vandenBosch1999,Jiang2008, McCavana2012,BoylanKolchin2008,Villalobos2013}. To go further, we also tried remaking the V$_\perp$ versus R$_{\rm{peri}}$ trend with a sample of objects whose mass ratios are restricted to be below 0.01, meaning dynamical friction will be weak. We found that the trend changes negligibly which further confirms that the effects of dynamical friction on the fit line can be considered strongly suppressed.

The result, shown in the lower-left panel is a reasonable match to the horizontal gradient shown by R$_{\rm{peri,All}}$ for mass ratios from $\sim10^{-4}$ up to 0.01 where most halos are found. This further supports that it is the dependency of V$_\perp$ on mass ratio that causes R$_{\rm{peri,All}}$ to also depend on mass ratio for the majority of the halos.

\begin{figure*}[ht]
\begin{center}
\includegraphics[width=16.8cm]{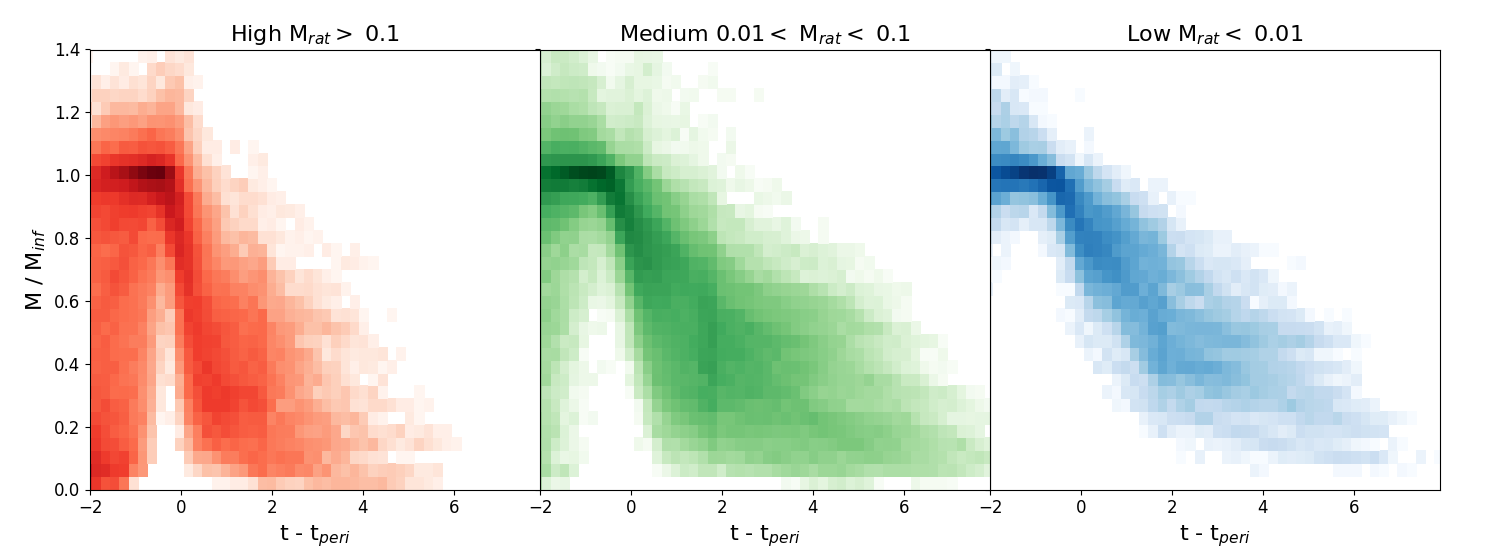}
\end{center}
\caption{Evolution of infaller halo masses (normalised by halo mass at infall time) plotted against time since pericentre passage in Gyr. Halos with higher mass ratios (M$_{\rm{rat}}$=M$_{\rm{inf}}$/M$_{\rm{host}}$) suffer more rapid tidal mass loss near pericentre, due to the dependency of V$_\perp$ on mass ratio, combined with dynamical friction for objects with M$_{\rm{rat}}>0.1$.}
\label{fig:MassLossMrat}
\end{figure*}

However, it can be seen that simply using the fitted R$_\perp$ dependency on V$_\perp$ does not reproduce the small median R$_{\rm{peri}}$ values where the mass ratio is $>0.1$ (i.e., the red pixels found on the right in the upper-left panel). At these mass ratios, dynamical friction begins to play a role as well (as we saw in the lower-left panel of Fig. \ref{fig:rperibecauseofvperp}), further reducing the R$_{\rm{peri}}$ value. In the lower-right panel, we attempt to isolate this effect by subtracting, pixel-by-pixel, the lower-left panel from the upper-left panel (R$_{\rm{peri,All}}$-R$_{\rm{peri,fit}}$). In this way, we can directly see how much smaller the R$_{\rm{peri}}$ values become when compared to a sample of halos in which dynamical friction is heavily suppressed due to being dominated by low mass ratio halos. We find that the additional dynamical friction at these higher mass ratios typically reduces R$_{\rm{peri,All}}$ by less than 10\% and only for infaller halos which are $>$5-10\% of their host mass, consistent with what we see in the lower-left panel of Fig. \ref{fig:rperibecauseofvperp}.

In summary, these results reveal that the dependency of V$_\perp$ on mass ratio drives R$_{\rm{peri,All}}$ to have the same dependency. When the mass ratio becomes very high ($>$5-10\%) dynamical friction further reduces R$_{\rm{peri,All}}$. Thus the mass ratio plays a key role on R$_{\rm{peri,All}}$. We note that in the top-left panel of Fig. \ref{fig:grids} there is not a clear gradient of R$_{\rm{peri,All}}$ in the vertical direction, meaning R$_{\rm{peri,All}}$ does not depend strongly on infall time. This is at first sight confusing as, when we split the sample by infall time, the R$_{\rm{peri}}$ distribution changed (see lower-right panel of Fig. \ref{fig:rperibecauseofvperp}). However, the black contours in the upper-left panel of Fig. \ref{fig:grids} show the halos are not distributed uniformly across the grid, and with earlier infalls the mass ratio becomes typically higher. Thus, it is in fact a dependency on mass ratio that drives the observed change in R$_{\rm{peri}}$ distributions with infall time.

Similarly, the dependency on infaller mass that is seen in the upper-right panel of Fig. \ref{fig:rperibecauseofvperp} can also be shown to be a result of the mass ratio. We tested this by separating the sample into a high and low infaller mass sample, and re-plotting them (see Appendix Fig. \ref{fig:infallsnap_mrat_grid_splitbyMhostMgal}). The choice of infaller mass causes us to sample different regions of the grid. For example, a high infaller mass subsample tends to fall in higher mass ratio pixels, which have small R$_{\rm{peri,All}}$ values. Thus, like with the infall time, there is little direct dependence of R$_{\rm{peri}}$ on infaller mass, except through its impact on the mass ratio, which is the more fundamental quantity dictating R$_{\rm{peri}}$.

In Fig. \ref{fig:MassLossMrat}, we directly demonstrate the impact of the mass ratio for tidal mass loss inside hosts. The three panels show the mass evolution of infaller halos, presented in the same way as in Fig. \ref{fig:tidalmassloss} (normalised at their infall time, against time since pericentre) for a high, medium and low mass ratio subsamples (see subtitles). By eye, the rate of tidal mass loss near pericentre is clearly a function of the mass ratio, due to R$_{\rm{peri}}$ depending on the mass ratio. We quantified this result by measuring the percentage of infaller halos that suffer strong tidal mass (defined as M/M$_{inf}<0.4$) by 2~Gyr after pericentre. For high, medium and low mass ratio infallers, 59.3\%, 37.5\% and 25.7\% suffer strong tidal mass loss, respectively. Thus, contrary to intuition, it is halos that are more massive with respect to their host that suffer more rapid tidal mass loss, due to infallers with a high mass ratio having reduced values of V$_\perp$, and additionally a further reduction in R$_{\rm{peri}}$ by dynamical friction when the mass ratio is $>0.1$. Thus, through a combination of both internal and external effects to the cluster, the mass ratio plays a key role on the efficiency of tidal mass loss in dense environments, with external effects dominating for low mass ratio infallers (M$_{\rm{rat}}<0.01$) that dominate by number.

\begin{figure*}[ht]
\begin{center}
\includegraphics[width=17cm]{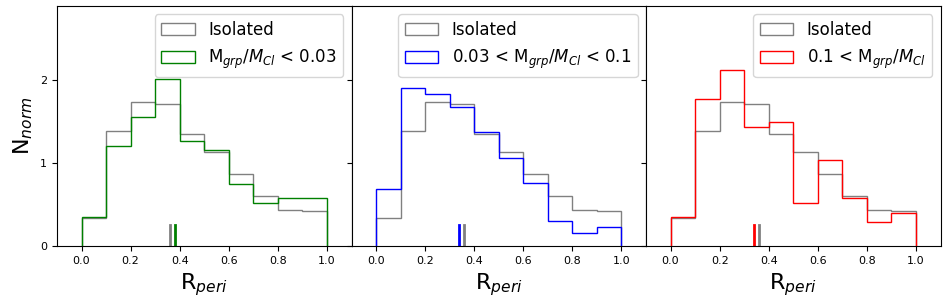}
\end{center}
\caption{R$_{\rm{peri}}$ distributions of all isolated infallers (grey histogram, same in all the panels for comparison), and infallers that are in a group when they enter the cluster. Left to right, the panels show substructures that have a low, medium and high mass ratio with the cluster. The small vertical line on the x-axis indicates the medians of the distributions.}
\label{fig:groupinfall}
\end{figure*}

\subsubsection{Group infallers}
 
In this section, we consider if halos that fall into cluster-mass hosts as members of groups have different R$_{\rm{peri}}$ distributions than those that fall into their hosts in isolation. When each infaller halo first crosses the R$_{200}$ of its host halo, we check to see if it is bound to any other halo that is more massive than the infaller halo (e.g., a group halo). If so, we label this halo as a `group infaller' and record the mass of the group halo (M$_{\rm{grp}}$). If not, we label this halo as an `isolated infaller'. We then record the resulting pericentre distances of the infaller halos and check to see if group membership alters their R$_{\rm{peri}}$ distributions significantly.

In Fig. \ref{fig:groupinfall}, we split our group infaller sample up by the group-to-cluster mass-ratio measured at the time of infall into the cluster (see panel legends from left to right). The distribution of the isolated infallers (shown in gray) is the same in all the panels for the purposes of comparison. All histograms are normalised by area. 

Overall, the R$_{\rm{peri}}$ distributions are  not strongly affected by group membership.  In \cite{Li2020} it was proposed that lower mass subhalos might have higher tangential velocities (and thus larger pericentres) as they are more easily perturbed by their proximity to the massive group halo. However, it seems there are likely two competing effects. It is true that halos that infall in groups have their own peculiar motions within the potential well of the group that may scatter their V$_\perp$ values. But counteracting this effect, dynamical friction acting on the group halo would tend to reduce R$_{\rm{peri}}$ for the group halo itself, and any group satellites that remain bound to the group halo could potentially be dragged down with it \citep{Choque2019}. In any case, we find that group membership does not play an an important role on the first pericentre distances of first infallers within clusters.

\begin{figure*}[ht]
\begin{center}
\includegraphics[width=8.5cm]{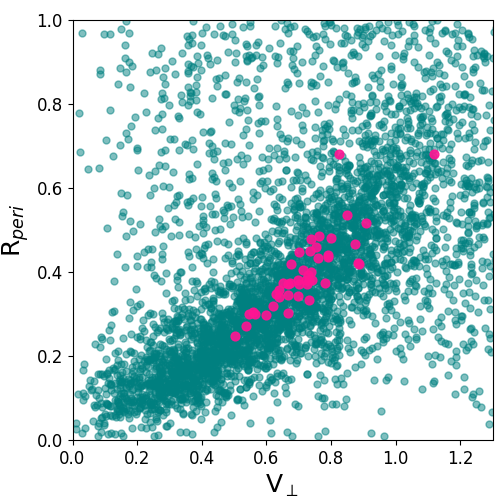}
\includegraphics[width=8.5cm]{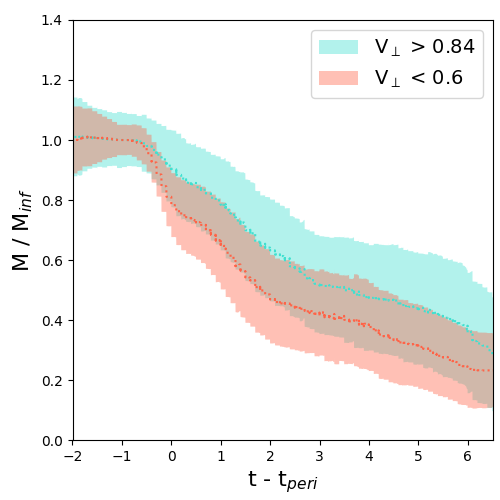}
\end{center}
\caption{Left: Cluster-to-cluster variations in the Rperi vs Vperp plot. Magenta symbols are the median R$_{\rm{peri}}$ and V$_\perp$ value for the first infallers into each individual cluster (i.e., one symbol per cluster). Teal symbols are all the individual first infallers into clusters (not split by cluster host). Right: Mass evolution of infallers, normalised by mass at infall time for the 6 clusters with the highest median V$_\perp$ value (red line; labelled `V$_\perp<$ 0.6'), and the 6 clusters with the lowest median V$_\perp$ value (blue line, labelled `V$_\perp$ $>$ 0.84'). The bold line is the median mass evolution track, and shading shows the standard deviation.}
\label{fig:clustertoclustervar}
\end{figure*}

\subsection{Cluster-to-cluster variations}
\label{sec:clustertoclustervar}

To test for variations between hosts, we split our sample of $z=0$ infallers into clusters by which host they fall into. For each individual host cluster, we calculate the median value of V$_\perp$ measured as they first cross the host's R200 and, after they reach pericentre, their median value of R$_{\rm{peri}}$. 

The magenta symbols in the left panel of Figure \ref{fig:clustertoclustervar} show the medians for the individual clusters, while the teal symbols show all the individual cluster infallers (i.e., not split by cluster host). It is clear that the median values of individual clusters follow a similar trend to the general trend of the individual infallers. Thus, between clusters, there is some variation in the typical V$_\perp$ value which in turn results in variation in their typical R$_{\rm{peri}}$ values.  

We confirmed that clusters with smaller median V$_\perp$ values indeed cause more rapid tidal mass loss. To do this, we select two subsamples of the clusters, each consisting of the 6 clusters with the highest and lowest median V$_\perp$ values. The mass evolution of the halos after falling into the cluster, normalised by their mass at infall time, is shown in the panel on the right. The clusters with typically lower V$_\perp$ (red line), and consequently smaller R$_{\rm{peri}}$, suffer stronger and more rapid tidal mass loss than those with larger V$_\perp$ (blue line). These results highlight that, even if clusters are similar in terms of their mass and/or internal tidal field, they may differ in their efficiency to cause tidal mass loss, and likely other environmental effects as well, as a result of differing peculiar motions inherited from the large scale surroundings.

This result also opens up the possibility that, in principle, it might be possible to observe the dynamics of first infallers in real clusters and gain insight into the strength of environmental effects acting near their centres. However, observationally measuring V$_\perp$ values is not trivial. Halos enter clusters with both a radial and tangential velocity component that may be difficult to separate cleanly with a single line-of-sight. We attempt to resolve this by searching for correlation between the median V$_\perp$ and the standard deviation of the cartesian velocity vectors V$_{\rm{x}}$, V$_{\rm{y}}$, and $V_{\rm{z}}$, measured in the cluster frame-of-reference at the moment of infall. We find a statistically significant correlation (Spearman's rank correlation coefficient $r_s$=0.40, $p$=0.005) for smaller standard deviations to result in lower V$_\perp$. However, practically speaking, the correlation is not sufficiently strong to cleanly distinguish between low and high V$_\perp$ clusters.

\begin{figure*}[ht]
\begin{center}
\includegraphics[width=17cm]{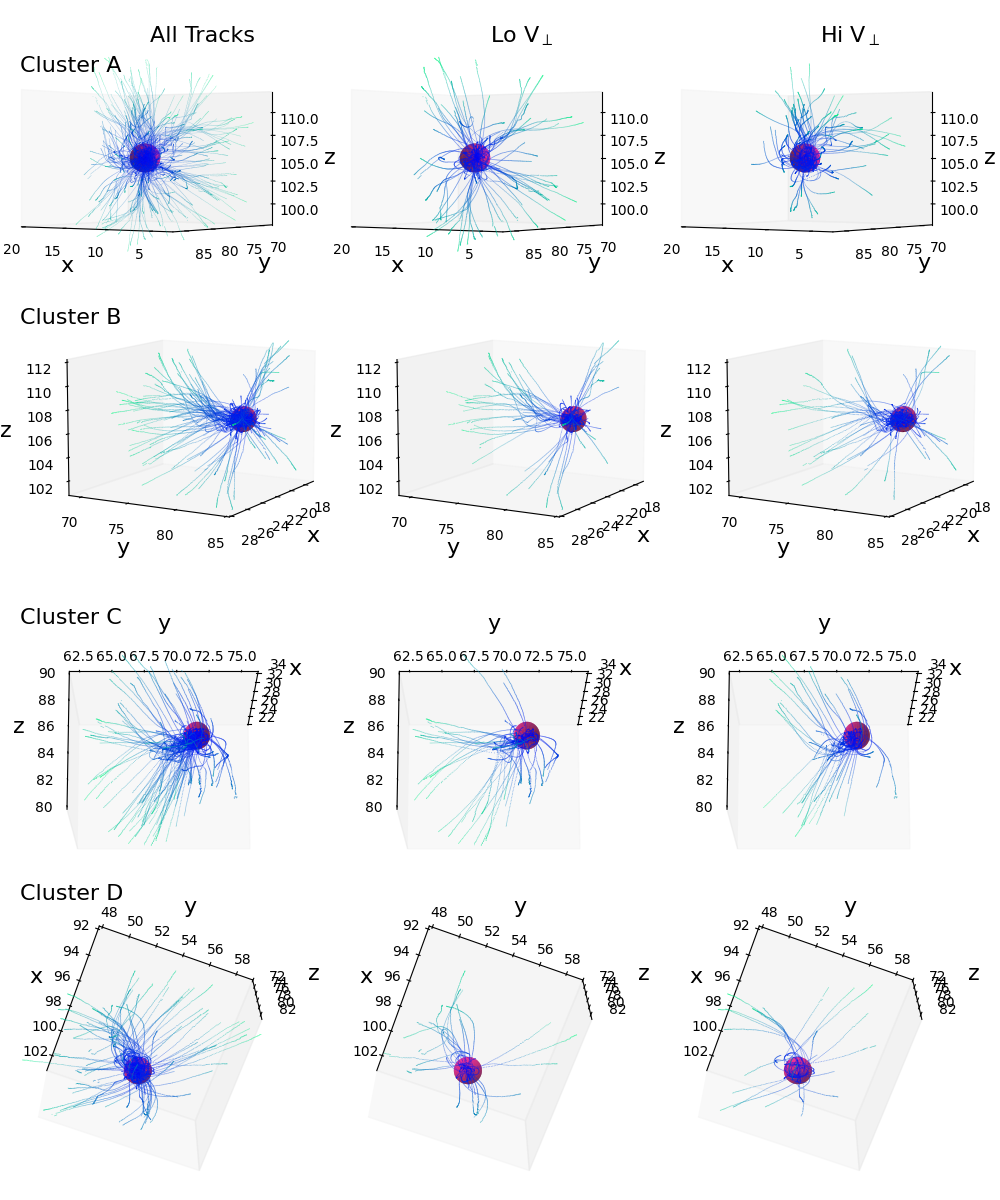}
\end{center}
\caption{Four representative examples of clusters (shown top to bottom; labelled `Cluster A--D'). A deep-pink sphere show the location and R$_{200}$ of the clusters at $z$=0. Blue lines show the trajectories of infaller halos, coloured darker blue according to their proximity to the cluster. The left column (labelled `All Tracks') shows the tracks of all the infallers and highlights the filamentary connections of each cluster. The middle and right column (labelled `Lo V$_\perp$' and `Hi V$_\perp$', respectively) presents the infaller tracks of low and high V$_\perp$ subsamples, with equal numbers shown for a fair visual comparison.}
\label{fig:filamentfeeding}
\end{figure*}

To add to the observational challenge, our median V$_\perp$ measurement is accumulated over time, as each halo enters the virial radius for the first time. However, at any instant in time there will be a mixture of first infallers and backsplash halos at R$_{200}$, and observationally it may be difficult to cleanly separate the two populations. We tried testing if the median V$_\perp$ measurement correlated with the velocity dispersion in the x, y, and z direction of all halos within a spherical shell of 1-2~R$_{200}$ and 2-3~R$_{200}$, when all are measured at a fixed time ($z=0$). But in this case we found no statistically significant correlation. We also tried correlating the median V$_\perp$ measurement with the average infall time of halos, and with the local number density in the outskirts of the clusters, but again found little correlation. Thus, while a correlation between median V$_\perp$ and R$_{\rm{peri}}$ could potentially provide a useful tool for understanding environmental effects in dense environments, some additional development is first required that is beyond the scope of this current study.

\subsection{Filament Feeding}
\label{sec:LSSfeeding}

So far, we have seen clear evidence for the important role played by V$_\perp$ in dictating the value of R$_{\rm{peri}}$ measurements of the majority of infaller halos, which in turn affects the amount of tidal mass loss they suffer too. In this section, we now consider how the large scale structure (LSS) surrounding the cluster can influence the measured V$_\perp$ values.

In Fig. \ref{fig:filamentfeeding}, we present four representative examples of clusters (shown top row to bottom; dark pink spheres are the $z=0$ cluster halo with radius of R$_{200}$). Blue lines show the trajectories of infaller halos, coloured darker blue according to their proximity to the cluster. The left panels show all the trajectories of infaller halos (column labelled `All Tracks'). From these trajectories, we produce a subsample of the lowest and highest V$_\perp$ values (columns labelled `Lo V$\perp$' and `Hi V$_\perp$', respectively), typically divided at a V$_\perp$ value of $\sim0.7$, but with small adjustments made to ensure the two subsamples are of equal size so they can be fairly compared visually.

In the `All Tracks' panels (left column), we can best see how the cluster is preferentially fed from particular directions. There is typically one dominant direction in these examples, but the second row (`Cluster B') clearly shows preferential feeding from three additional more minor filaments, and the top row (`Cluster A') shows at least five equally significant filaments are feeding the cluster. It is notable that if we compare the low and high V$_\perp$ panels (centre and right columns), the primary filaments feeding the clusters are more clearly traced out by the low V$_\perp$ panels. This means that halos with low V$_\perp$ values preferentially enter the cluster along filaments that feed the cluster. In comparison, the high V$_\perp$ trajectories tend to join the filaments or cluster from directions that are more perpendicular to the filaments -- see for example the high V$_\perp$ panel of the third row (`Cluster C'). These trajectories that join the filaments perpendicularly are reminiscent of halos moving along walls before joining the filaments.

However, we note that the results are not clear cut. The trajectories in Fig. \ref{fig:filamentfeeding} show that some halos that are not associated with filaments have low V$_\perp$ orbits. And, often halos associated with filaments enter their cluster with high V$_\perp$ values. Therefore, whether the halos are in filaments is not the only deciding factor. This may partly be because some halos that are located in filaments have not yet relaxed into the filament potential and are currently caught in the process of crossing the filament \citep{Jhee2022}. Therefore, their V$_\perp$ values may be quite high despite being located in the filament. Only the fraction of the halos that are well virialised in the filament may present the low V$_\perp$ values. Thus, in this case, the presence of a filament simply increases the chances for some halos to enter the host halo along the filament conduit with reduced V$_\perp$ values (and consequently to have reduced R$_{\rm{peri}}$, and suffer more significant environmental effects within the host halo), but it does not guarantee this. Also, an additional factor will be the large scale tidal field in which the filaments are situated which is expected to also impact on the V$_\perp$ values both inside filaments and when they are accreted onto clusters \citep{Shi2015}. 

In any case, we do see a clear dependency of V$_\perp$ on infaller-to-host halo mass-ratio in the upper-right panel of Fig. \ref{fig:grids} which filaments likely play an important part in producing. It is well known that filaments are mass segregated in that more massive halos are preferentially found closer to filaments \citep{Kraljic2018,Libeskind2018}. As a result, the dynamics of halos around filaments may partly drive the clear connection between the mass ratio and V$_\perp$ that we detect in our simulations.

\section{Summary and Conclusions}
\label{sec:conclusions}

Using dark matter-only cosmological simulations, we study the trajectories of dark matter halos that fall into group- and cluster-mass host halos for the first time (referred to as `infaller halos'). We measure their first pericentre distance (R$_{\rm{peri}}$, the minimum distance to the cluster centre normalised by the cluster R$_{200}$), and find that the strength of environmental effects within the host halo are strongly influenced by their R$_{\rm{peri}}$ value. We attempt to determine what parameters dictate the resulting R$_{\rm{peri}}$ value. Our main results are summarised as follows:

\begin{enumerate}
    \item The primary factor influencing the R$_{\rm{peri}}$ value of most infaller halos is a measure of their peculiar motion, V$_\perp$, the perpendicular component of the orbital velocity, which we measure at the moment the infaller halo enters their host halo for the first time. Higher V$_\perp$ results in larger R$_{\rm{peri}}$ which means the strength of tidal mass loss to be dictated by the dynamics of the surrounding large scale structure. 
    \item Dynamical friction can also play a role in decaying the infaller halo's orbital energy and thus reducing R$_{\rm{peri}}$. But it has limited time to act between the infaller first entering the group/cluster and reaching the first pericentre, and we find it is only effective on the 5\% of the infaller halos with masses $\gtrsim 10$\% than their host mass. 
    
    \item  Halos in filaments have increased chances of having a smaller V$_\perp$ value but it isn't guaranteed, due to the complex and multi-scale velocity field of the large scale structure. However, we uncover a tendency for infaller halos with a high mass ratio to have reduced V$_\perp$. As a result, the mass ratio is shown to have influence over the first infallers orbit and its resulting tidal mass loss. Somewhat counter-intuitively, higher mass satellites are destroyed more rapidly than lower mass satellites, for a fixed host mass, as a result of this.    
    \item From cluster to cluster, there are significant variations in their median R$_{\rm{peri}}$ values, directly as a result of differing median V$_\perp$ values. Thus, even similar mass clusters (with similar internal tidal fields) subject their first infallers to differing levels of tidal stripping (and other environmental mechanisms) at their cores, as a result of factors that are entirely external to the clusters.    
    
\end{enumerate}

These results highlight how the efficiency of environmental mechanisms occurring at the centres of groups and clusters can be significantly altered by the peculiar velocities of halos inherited from global dynamics within the large scale structure, far beyond the cluster virial radius. This emphasises the importance of considering both the internal properties of dense environments and their large scale surroundings in order to better understand the efficiency of environmental transformation mechanisms.

\section*{Acknowledgements}
We thank the anonymous referee for the detailed report that helped improve the manuscript. We are very grateful to Stephanie Tonnesen and Sean McGee for comments and helpful discussions on an earlier version of this paper. This research was supported by the Korea Astronomy and Space Science Institute under the R\&D program (Project No. 2022-1-830-05), supervised by the Ministry of Science and ICT. J.S. acknowledges support from the National Research Foundation of Korea grant (2021R1C1C1003785) funded by the Ministry of Science, ICT \& Future Planning.


\onecolumngrid{
    \bibliographystyle{apj}
    \bibliography{bibfile}}

\appendix
\counterwithin{figure}{section}

\section{Infall Time -- Mass Ratio grids split by host and infaller mass}

\begin{figure*}[hb]
\begin{center}
\includegraphics[width=17cm]{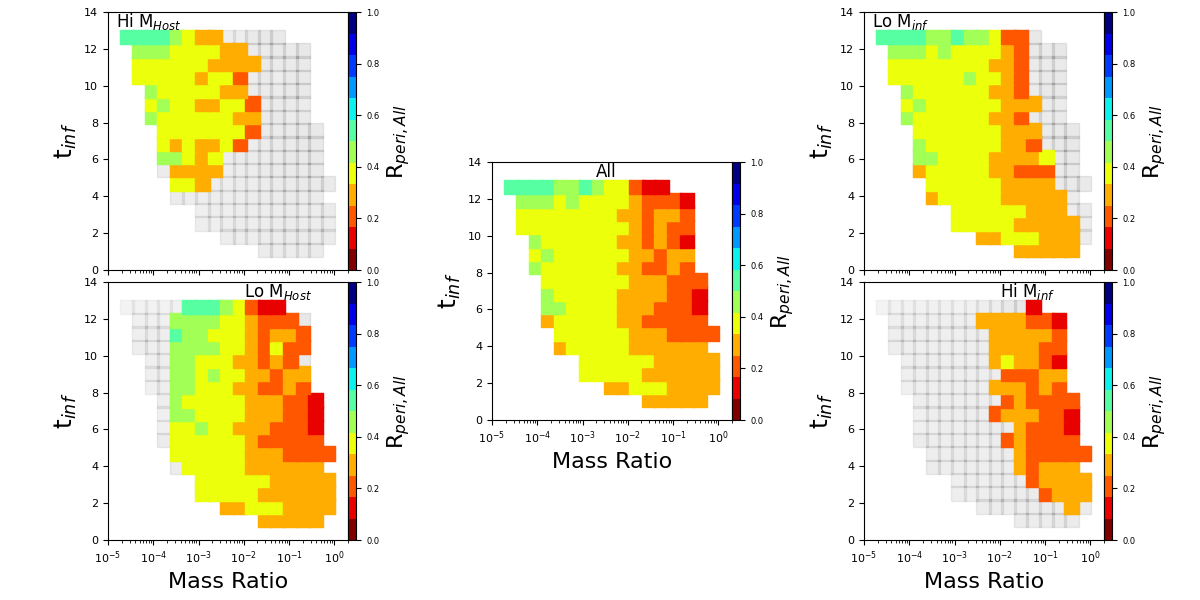}
\end{center}
\caption{Age of the Universe in Gyr at the time of infall (t$_{inf}$) versus Mass Ratio grids, coloured by the median R$_{\rm{peri}}$ of all halos in a pixel (R$_{\rm{peri,All}}$). In the central panel, the total sample is shown (labelled `All'). In the left panels, we split the sample by host mass (labelled `Hi M$_{\rm{host}}$' and `Lo M$_{\rm{host}}$'). In the right panels, we split the sample by infaller mass (labelled `Lo M$_{\rm{inf}}$' and `Hi M$_{\rm{inf}}$').}
\label{fig:infallsnap_mrat_grid_splitbyMhostMgal}
\end{figure*}

In the centre panel of Figure \ref{fig:infallsnap_mrat_grid_splitbyMhostMgal}, we plot a grid of the age of the Universe in Gyr at the time of infall versus Mass Ratio for the full sample (labelled `All'), coloured by median R$_{\rm{peri}}$ (labelled as R$_{\rm{peri,All}}$ on the colour bar). This is the same as shown in the top-left panel of Fig. \ref{fig:grids}. However, in the surrounding panels we show how the grids appear if we select subsamples of the `All' sample based on their host mass (left panels) or their infall halo mass (right panels). Infallers into high and low host-mass halos (labelled `Hi M$_{\rm{host}}$' and `Lo M$_{\rm{host}}$', respectively) are divided at M$_{\rm{host}}=10^{13.5}$~M$_\odot$.  High and low mass infallers halos (labelled `Hi M$_{\rm{inf}}$' and `Lo M$_{\rm{inf}}$', respectively) are divided at M$_{\rm{inf}}=10^{11.5}$~M$_\odot$. To reduce noise, we only colour pixels with at least 20 halos. 

The main result to note is that, overall, the R$_{\rm{peri,All}}$ values do not change substantially when we subsample by host mass and infall halo mass. I.e., in the regions where the pixels are coloured, the colours are quite similar to that shown in the `All' panel. This indicates that the host mass and infaller mass do not independently have a strong impact on R$_{\rm{peri}}$ when the mass ratio is fixed. In other words, R$_{\rm{peri,All}}$ does not strongly depend on host mass or infaller mass. However, the distributions of R$_{\rm{peri}}$ change with these parameters (as seen in Fig. \ref{fig:rperibecauseofvperp}) because when we subsample by these parameters, different regions of the grid are included in the R$_{\rm{peri}}$ distribution. For example, in the lower-right panel, we see that high mass infallers tend to be high mass ratio objects, where the pixels tend to be more red, indicating smaller R$_{\rm{peri,All}}$ values.

\newpage

\section{Time Resolution for accurate R$_{\rm{peri}}$ measurements}

\begin{figure*}[hb]
\begin{center}
\includegraphics[width=8.9cm]{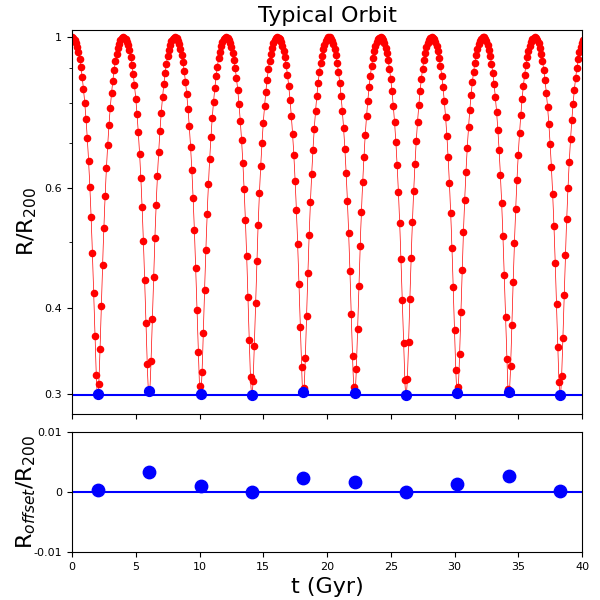}
\includegraphics[width=8.9cm]{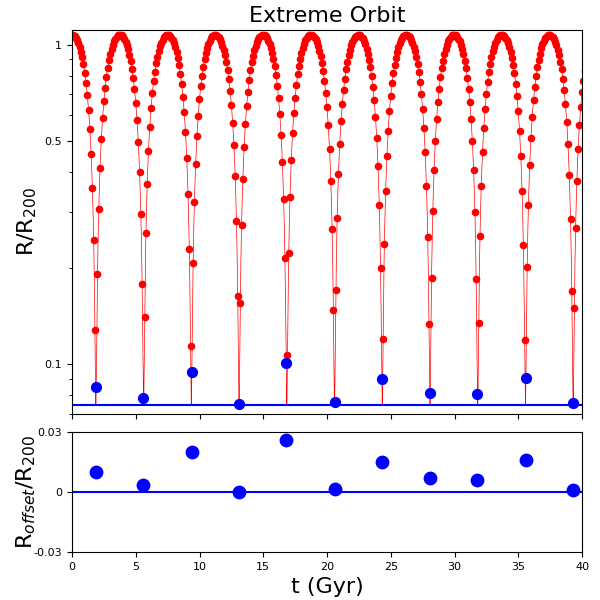}
\caption{Evolution of R$_{\rm{peri}}$ with time for a point mass orbiting in a cluster-like fixed analytical potential for a typical orbit of a satellite in a cluster (left) and for a more extreme plunging orbit (right). The red line shows the R$_{\rm{peri}}$ evolution for 1~Myr timesteps, while the red symbols show the R$_{\rm{peri}}$ evolution with 100~Myr timesteps, similar to the cosmological simulation. Blue symbols show the minimum R$_{\rm{peri}}$ values seen with the 100~Myr timesteps. The horizontal line shows the true R$_{\rm{peri}}$ value. With 100~Myr timesteps, typically the recovered R$_{\rm{peri}}$ measurement matches the true measurement to within 1\%.}
\label{fig:missedperis}
\end{center}
\end{figure*}

Typically, the snapshots of our cosmological simulation are approximately 100~Myr apart. In order to assess the accuracy of our R$_{\rm{peri}}$ measurements given that the true pericentre could occur between the snapshots, we conduct a numerical experiment. We used a simple particle integrator to track the orbital evolution of a point mass within the gravitational potential well of a cluster-mass dark matter halo. 

We consider two possible orbits; (left panels) a typical orbit for first infallers and (right) a more extreme plunging infall. For the typical orbit (left panels), we place the particle on a plunging orbit that travels from the edge of the cluster (R$_{200}$) to 0.3 R$_{200}$, which is a fairly typical value (see Fig \ref{fig:mainrperidistrib}). For the cluster potential we chose an NFW \citep{Navarro1996} potential with M$_{200}$=1.1$\times$10$^{14}$~M$_\odot$, and a typical concentration parameter $c=6.6$ for halos of this mass. We conduct the simulation for several thousand Gigayears, resulting in multiple pericentre passages from which to measure the accuracy of the pericentre measurements.

The smooth red line in the top panels of Figure \ref{fig:missedperis} shows the radial evolution (normalised by R$_{200}$) of the point mass with 1~Myr timesteps (showing only the first 40~Gyr of evolution). The blue horizontal line is the true pericentre of the orbit. The red filled circles show the radius when only seen every 100~Myr, like in the cosmological simulation. The blue filled symbols show the closest approach to the true pericentre with the 100~Myr time-sampling. The closeness of the blue symbols to the blue horizontal line indicate that the R$_{\rm{peri}}$ measurements with 100~Myr time-sampling are very similar to their true pericentre value. The lower panels show that the difference is less than 1\% of R$_{200}$.

In the more extreme infaller case (right panels), we select objects with the smallest pericenters and highest pericentre velocities from the right panel of Fig. \ref{fig:tidalmassloss}. We then place a point mass on an extreme orbit (V$_{\rm{peri}}$=2.2 and V$_{\rm{peri}}$=0.07) within a more massive cluster (3$\times$10$^{14}$~M$_\odot$) and repeat the experiment. We find that, even in this more extreme case, over half of the pericentre passages are measured to within 1\% of the true value, and the worst case is only 2.8\% which is generally more precise than the width of the bars in our histogram results. This demonstrates that our R$_{\rm{peri}}$ measurements are not strongly affected by the time resolution of our cosmological simulations.




\end{document}